\documentclass[aps,prd,onecolumn,nofootinbib,preprint,superscriptaddress]{revtex4-1}
\usepackage{hyperref}
\usepackage{ulem}
\usepackage{graphicx}				
\usepackage{amssymb}
\usepackage{amsfonts,amsmath,amssymb,graphics,epsfig}
\usepackage{comment}
\usepackage{footnote}
\usepackage{graphicx}
\usepackage{dcolumn}
\usepackage{bm}

\usepackage{geometry}                		
\usepackage{graphicx}				
\usepackage{amssymb}
\usepackage[T1]{fontenc} 
\usepackage{graphicx}
\usepackage{graphicx}
\usepackage{subcaption}
\usepackage{caption}
\usepackage{mwe}
\usepackage{amsmath}
\usepackage{slashed}
\usepackage{young}
\usepackage{amssymb}
\usepackage[usenames,dvipsnames]{color}
\usepackage{amsmath}
\usepackage{slashed}
\usepackage{young}
\usepackage{amsfonts}
\usepackage{hyperref}
\usepackage{amssymb}
\usepackage{footnote}
\usepackage[usenames,dvipsnames]{color}
\usepackage{natbib}

\begin{document}
\title{Tunneling Between Schwarzschild-de Sitter Vacua}
\author{Klaountia Pasmatsiou}%
 \email{kxp265@case.edu}
\affiliation{%
Physics Department/CERCA/ISO \\ 
Case Western Reserve University Cleveland, Ohio 44106-7079, USA}%

\date{\today}

\begin{abstract}
We extend the study of the effect of static primordial black holes on vacuum decay. In particular, we compare the tunneling rates between vacua of different values of the cosmological constant and black hole mass by pointing out the dominant processes based on a numerical examination of the thin wall instanton. Three distinct cases are considered, namely the nucleation of a true vacuum bubble into the false vacuum, the nucleation of a false vacuum bubble into the true vacuum as well as the Farhi-Guth-Guven mechanism. As a proof of concept, it is shown that in order to increase the transition rate into an inflating region, we find that not only is the inclusion of a black hole necessary, but the inclusion of a cosmological constant in the initial phase is also required. Among the cases studied, we show that the most likely scenario is the elimination of inhomogeneities in the final phase.
\end{abstract}

\maketitle

\newpage

\section{Introduction}

Research on false vacuum decay in quantum field theory was prompted by the work of Sydney Coleman et al. \cite{Coleman:1977py, Callan:1977pt, Coleman:1980aw}. The effect of gravitation on bubble nucleation has sparked intense study over the last forty years, leading to interesting investigations such as the effect of black holes on the nucleation rates of true vacuum bubbles. Early work on the topic can be found in \cite{Hiscock:1987hn, Berezin:1987ea, Moss:1984zf}, while more recent developments, motivated by the role of impurities in the decay rates of first order phase transitions, are addressed in \cite{Gregory:2013hja,Burda:2015yfa}. In the latter, it was shown that by relaxing the initial assumption of homogeneity of de-Sitter spacetime, the inclusion of black holes, as seeds of inhomogeneity, leads to enhanced decay rates. As a result, this process could affect the lifetime of the Higgs vacuum \cite{Gregory:2016xix, Burda:2015isa}, increasing dramatically the probability of vacuum decay. 

In this paper, we consider the Euclidean instanton approach\footnote{For an alternative, we refer the reader to the Lorentzian WKB approach \cite{Farhi:1989yr, Fischler:1990pk, Fischler:1989se}}. While this approach has been applied in the context of false vacuum decay (downward tunneling) \cite{Gregory:2013hja, Burda:2015yfa}, the present work extends the analysis to the nucleation of false vacuum bubbles within a low-energy true vacuum (upward tunneling). Our aim is to explore the parameter space and compare the tunneling rates between the initial and final states with a cosmological constant and/or a black hole to the standard upward tunneling scenario.  The latter has been explored in \cite{Lee:1987qc} while false vacuum bubbles of a de Sitter (dS) interior and a Schwarzschild-de Sitter (SdS) exterior were discussed in \cite{Aguirre:2005xs,Aguirre:2005nt}. We present the most general expression for the tunneling rates of false and true vacuum bubbles with a SdS interior/exterior to determine which processes are favorable due the inclusion of a cosmological constant in the initial phase.

A case of particular interest is that of the Farhi-Guth-Guven (FGG) mechanism for understanding the nucleation of inflating regions (false vacuum bubbles) from non inflating ones \cite{Farhi:1989yr, Aguirre:2005nt, Aguirre:2005xs}. A proper understanding of this mechanism could shed light on the beginning of inflation\cite{Guth:1980zm}. In this work, we apply the Euclidean approach to the FGG mechanism as well and we present a relative comparison between the FGG mechanism and the tunneling upward in the potential with a nonzero cosmological constant in both vacua.  Although we focus on vacua with a positive cosmological constant, we note that work has been done on the bubble nucleation in Schwarzschild-Anti de Sitter (S-AdS) spacetimes \cite{Burda:2015yfa} and a possible implication on the information loss problem can be found in \cite{Chen:2015lbp, Chen:2015ibc, Sasaki:2014spa}.

The manuscript is organized as follows: In section \ref{2} we provide the formalism of constructing the thin wall instanton. In section \ref{3}, we perform the numerical examination of the most general instanton. Finally, in section \ref{4} we discuss the FGG mechanism in the context of conical singularities. After discussing our results in section \ref{5} we provide the reader with details about conical angles in Appendix A.

\section{Constructing the instanton}\label{2}

We start our discussion, by reviewing the formalism presented in \cite{Gregory:2013hja,Burda:2015yfa}. We consider two Schwarzschild-de Sitter spacetimes with arbitrary cosmological constants separated by a thin wall of constant tension. By performing the Wick rotation $t \rightarrow - i \tau$, the metric on each side of the wall reads,
\begin{align} \label{eq:1}
ds^2= f(r) d\tau_{\pm}^2 + \frac{dr^2}{ f(r)} + r^2 d\Omega^2,
\end{align}
where $f(r)=1- \frac{2 G M_{\pm}}{r} - \frac{\Lambda_{\pm}r^2}{3}$. Here, $M_{+}$ is the mass of the black hole outside the bubble while $M_{-}$ is the mass of the remnant black hole inside the bubble. Moreover, $\Lambda_{+}$  and $\Lambda_{-}$ are the exterior and interior cosmological constants, respectively. The wall is parametrized by $r=R(\lambda)$ and the Israel junction conditions \cite{nuovocimento} lead to,
\begin{align}\label{eq:3}
f_{+} \dot{\tau}_{+} - f_{-} \dot{\tau}_{-}= - 4 \pi G \sigma R.
\end{align}
The radius of the bubble, $R$, is a function of the proper time $\lambda$, the dot represents the derivative with respect to $\lambda$, and $\sigma$ is the tension of the wall. Using \footnote{This relation comes from the fact that the induced metric must be the same on both sides of the wall parametrized by $\lambda$.},
\begin{equation}\label{eq:4}
f_{\pm} \dot{\tau}_{\pm} + \frac{\dot{R}^2}{f_{\pm}}=1,
\end{equation}
and (\ref{eq:3}) we arrive at the equation which describes the evolution of the bubble wall,
\begin{equation}\label{56}
\dot{R}^2= 1- \frac{2 G M_{-} }{R}- \left(\frac{R}{\gamma}+\frac{G \Delta M}{2 \bar{\sigma} R^2}\right)^2- \frac{R^2}{l^2_{-}},
\end{equation} 
where
 \begin{align}
& \gamma\equiv\frac{4 \bar{\sigma} \mu^2 }{1+ 4  \bar{\sigma}^2 \mu^2}, & \frac{1}{\mu^2}\equiv \frac{1}{l_{+}^2}-\frac{1}{l_{-}^2}.
 \end{align}
Here, $l_+$ ($l_-$) is the dS length inside(outside) the bubble, $ \Delta M=M_{+}-M_{-}$ and $\bar{\sigma}=2 \pi G {\sigma}$. The term $ \frac{R^2}{l^2_{-}}$ arises from the nonzero value of the cosmological constant in the true vacuum. 
Combining (\ref{eq:4}), (\ref{56}), the evolution of the time coordinate is given by 
\begin{equation}\label{eq:9}
f_{\pm} \dot{\tau}_{\pm}=\mp\bar{\sigma} R -\frac{\Delta f}{4 \bar{\sigma} R }.
\end{equation}

\subsection{Tunneling to lower values of the cosmological constant}\label{2.1}
Generally, the bubble nucleation rate reads,
\begin{equation}\label{229}
\Gamma=A e^{-\frac{B}{ \hbar}},
\end{equation}
where $B$ is the ``bounce'' and $A$ is a prefactor. It describes the probability to penetrate and escape a potential barrier \footnote{For the rest of the paper, we ignore $A$, we set $\hbar=1$ and we calculate the bounce.}. We begin by considering tunneling from a higher value of the cosmological constant to a lower one while we assume the existence of a Schwarzschild black hole both in the initial and final state. The Euclidean action for this case, is given by  \cite{Gregory:2013hja,Burda:2015yfa},
 \begin{eqnarray}\label{eq:1908}
I_{E}&=I_{M_{+}} + I_{M_{-}} + I_{\mathcal{W}}=\nonumber\\&&=-\frac{1}{16 \pi G}\int_{M_{+}}\sqrt{g}(R_{+}-2 \Lambda_{+})-\frac{1}{16 \pi G}\int_{M_{-}}\sqrt{g}(R_{-}-2 \Lambda_{-})\nonumber\\&&+\frac{1}{8\pi G}\int_{\partial M_{+}}\sqrt{h}K_{+}
-\frac{1}{8\pi G}\int_{\partial M_{-}}\sqrt{h}K_{-}+\int_{W}\sigma \sqrt{h},
\end{eqnarray}
where $R$ is the Ricci scalar, $K$ is the trace of the extrinsic curvature, $\Lambda$ is the cosmological constant and $\sigma$ is the surface tension of the bubble wall. The subscript + and - denotes the outside and inside region of the wall, respectively. To proceed, we need to explicitly calculate the Euclidean action on each side of the wall.

Before doing so, it is useful to mention that the issue of conical singularities has been explored in \cite{Gregory:2013hja, Fursaev:1995ef}, and the integral over the Ricci scalar for near horizon geometries, takes the form,
\begin{equation}
\int d^4x \sqrt{g} R \sim 4 \pi \Delta A,
\end{equation}
while the Gibbons-Hawking boundary term reads,
\begin{equation}
\int d^3x \sqrt{h} K \sim -2  \pi (1-\Delta) A,
\end{equation}
where $\Delta$ represents the deficit angle and $A$ is the area of the conical defect. Combining these terms together, we are left with an action that doesn't depend on the conical deficit \cite{Gregory:2013hja},
\begin{equation}
I\sim-\frac{A}{4G}.
\end{equation}
Let us write the expressions for the Euclidean action and from (\ref{eq:1908}), we have three distinct contributions to consider, as derived in \cite{Burda:2015yfa} :
\begin{itemize}  
\item Outside the bubble ($M_+$)
The Euclidean action for the exterior of the bubble is
 \begin{align}\label{eq:250}
 I_{M_{+}}=-\frac{A_{c_+}}{4 G} +\frac{\beta}{4 G} \left(\frac{A_{c_+}}{\beta_{c_+}}-\frac{2 \Lambda_{+} r_{{c_+}}^3}{3} + 2 G M_{+}\right)-\frac{1}{4 G}\int d\lambda R^2 f'_{+}\dot{\tau}_{+},
\end{align}
where $A_{c_+}= 4 \pi r_{c_+}^2$ represents the area of the cosmological horizon and $\beta$ is the periodicity of $\tau$ and in general is different than the periodicity of the cosmological horizon $r_c$, $\beta_{c_+}$ (See Appendix \ref{6} for a pedagogical discussion on conical angles). Using,
\begin{align}
R^2 f'_{+}&= 2 G M_{+}-\frac{2 \Lambda_{+}R^3}{3}, &\beta_{c_+}&=-\frac{4 \pi {r_{c_+}}^2}{2 G M_{+}-\frac{2 \Lambda_{+} {r_{c_+}}^3 }{3}},
\end{align}
where the prime represents the derivative with respect to $R$, the term in the parenthesis becomes zero and the action is independent of the conical angle.
\item Inside the bubble ($M_-$) 

The Euclidean action for the interior of the bubble is
\begin{equation}\label{eq:252}
I_{M_{-}}=-\frac{A_{-}}{4 G} +\frac{\beta}{4 G} \left(\frac{A_{-}}{\beta_{h _{-}}}+\frac{2 \Lambda_{-} r_{h_{-}}^3}{3} - 2 G M_{-}\right)+\frac{1}{4 G}\int d\lambda R^2 f'_{-}\dot{\tau}_{-},
\end{equation}
where $A_{-}$ represents the area of the black hole horizon and $\beta$ is the periodicity of $\tau$ and is different than the periodicity of the black hole horizon $r_{h_{-}}$, $\beta_{h _{-}}$. Using,
\begin{align}
R^2 f'_{-}&= 2 G M_{-}-\frac{2 \Lambda_{-}R^3}{3},&\beta_{h_{-}}&=-\frac{4 \pi r_{h_{-}}}{1-\Lambda_{-} r_{h_{-}}^2 },
\end{align}
we, again, see that the result does not depend on the conical angle.
\item Bubble wall ($\mathcal{W}$)

The action for the wall is
\begin{eqnarray}\label{eq:254}
I_{\mathcal{W}} &&=\frac{1}{8\pi G}\int_{\partial M_{+}}\sqrt{h}K_{+}-\frac{1}{8\pi G}\int_{\partial M_{-}}\sqrt{h}K_{-}+\int_{\mathcal{W}}\sigma \sqrt{h}\nonumber\\&&=\frac{1}{2 G} \int{d\lambda R(f_{+}\dot{\tau}_{+}-f_{-}\dot{\tau}_{-})}.
\end{eqnarray}
\end{itemize}
We have included the Gibbons-Hawking boundary terms induced by the wall  and we made use of the Israel junction conditions which can be written as, $K_{+} - K_{-}=- 12 \pi G \sigma$. Combining (\ref{eq:250}), (\ref{eq:252}) and (\ref{eq:254}), we arrive at
\begin{equation}
I_{E}=-\frac{A_{-}}{4 G} -\frac{A_{c_+}}{4 G} +\frac{1}{4 G} \int d\lambda [(2 R f_{+}-R^2  f'_{+})\dot{\tau}_{+}-(2 R f_{-}-R^2  f'_{-})\dot{\tau}_{-}].
\end{equation}
The ``bounce'' is obtained by subtracting the background Euclidean action from the Euclidean action for the bubble wall solution \cite{Burda:2015yfa},
\begin{equation}\label{209}
B_{down}=I_{E} -I_{SdS_+}=\frac{A_{+}}{4 G} -\frac{A_{-}}{4 G} +\frac{1}{4 G} \int d\lambda [(2 R f_{+}-R^2  f'_{+})\dot{\tau}_{+}-(2 R f_{-}-R^2  f'_{-})\dot{\tau}_{-}],
\end{equation}
where
\begin{equation}\label{20034}
I_{SdS_+}=-\frac{A_{c_+}}{4 G} -\frac{A_{+}}{4 G}.
\end{equation}
In (\ref{20034}), $A_{+}$ represents the area black hole horizon.

\subsection{Limiting case of no black hole}\label{2.2}

For future reference, we give the expression for the bounce of the dS-dS transition \cite{Burda:2015yfa},
\begin{equation}\label{eq:123}
B^{down}_{M_{\pm}=0}=2 \int{d\lambda R(\dot{t}_{+}-\dot{t}_{-})},
\end{equation}where $B^{down}_{M_{\pm}=0}$ corresponds to the Coleman-DeLuccia bounce $(B_{CDL})$ for downward tunneling. We solve the equation for the bubble wall (\ref{56}) with $ R[-\gamma \pi/(2 \sqrt{1 + \zeta})] =0$ as the initial condition to obtain,
\begin{equation}\label{eq:121}
R[\lambda] =\frac{\gamma \cos \left[
  \frac{ \sqrt{1 + \zeta} \lambda}{\gamma}\right]}{\sqrt{1 + \zeta}},
\end{equation}
where $\zeta\equiv\frac{\gamma^2}{l^2_{-}}$. The equations which describe the evolution of the time coordinate are,
\begin{align}\label{eq:122}
&\dot{t}_{+}=\frac{(1- 2 \bar{\sigma} \gamma) \cos \left[
  \frac{  \sqrt{1 + \zeta} \lambda}{\gamma}\right]}{\sqrt{1 + \zeta}\left(1-\frac{\gamma^2\cos\left[
  \frac{  \sqrt{1 + \zeta} \lambda}{\gamma}\right]^2}{(1+\zeta) l_{+}^2}\right)}, & \dot{t}_{-} = \frac{ \cos\left[
  \frac{  \sqrt{1 + \zeta} \lambda}{\gamma}\right]}{\sqrt{1 + \zeta}\left(1-\frac{\zeta \cos\left[
  \frac{  \sqrt{1 + \zeta} \lambda}{\gamma}\right]^2}{(1+\zeta) }\right)}.
  \end{align}
Plugging (\ref{eq:121}) and (\ref{eq:122}) into (\ref{eq:123}) leads to
\begin{equation}\label{eq:500}
B^{down}_{M_{\pm}=0}=\frac{\pi  \left[l_{+}^4 \left(4 \bar{\sigma} ^2 l_{-}^2+1\right)+l_{+}^2 \left(d+4 \bar{\sigma}
   ^2 l_{-}^4-2 l_{-}^2\right)-l_{-}^2 d+l_{-}^4\right]}{2 G d},
\end{equation}
where $d= \sqrt{l_{+}^4 \left[4 \bar{\sigma} ^2 l_{-}^2+1\right)^2+l_{+}^2 \left(8 \bar{\sigma} ^2 l_{-}^4-2 l_{-}^2\right]+l_{-}^4}$. This result agrees with \cite{Burda:2015yfa,Parke:1982pm}.

\subsection{Tunneling to higher values of the cosmological constant}\label{2.3}

Having set the basis so far, we proceed in presenting the novel part of our work by deriving the general expression for the tunneling rate between a spacetime of a lower value to a higher value of the cosmological constant \footnote{This is a natural generalization of a similar expression given in \cite{Lee:1987qc}, eq. (4.3). There, $I_{f}$ corresponds to $I_{dS_+}$ while $I_{t}$ corresponds to  $I_{dS_-}$.},
\begin{equation}
\frac{\Gamma_{down}}{\Gamma^{up}}=\frac{e^{-I_{E}+I_{SdS_{+}}}}{e^{-I_{E}+I_{SdS_{-}}}}=e^{I_{SdS_{+}}-I_{SdS_{-}}},
\end{equation}
or using (\ref{229}),
\begin{equation}\label{eq:100}
B^{up}={I_{SdS_+}-I_{SdS_-}}+ B_{down}.
\end{equation}
Using (\ref{209}), (\ref{20034}) as well as,
\begin{equation}\label{bb}
I_{SdS_-}=-\frac{A_{c_-}}{4 G} -\frac{A_{-}}{4 G},
\end{equation}
we obtain,
\begin{equation} \label{219}
B^{up}=-\frac{A_{c+}}{4 G}  +\frac{A_{c-}}{4 G}  +\frac{1}{4 G} \int d\lambda [(2 R f_{+}-R^2  f'_{+})\dot{\tau}_{+}-(2 R f_{-}-R^2  f'_{-})\dot{\tau}_{-}].
\end{equation}

Even though in flat spacetime, energy conservation forbids the upward tunneling at zero temperature, if the true vacuum is a de Sitter space-time, and when the temperature is nonzero, thermal fluctuations allow the creation of such a bubble \cite{Lee:1987qc}. Given that this is a possible process\footnote{One could consider the Hawking-Moss instanton as a competing process for example, and it would interesting to see which one dominates the transition rate in a future work.}, we see its generalization for black holes. The zero-mass limit of (\ref{219}) reads
\begin{equation}\label{919}
B^{up}_{M_{\pm}=0}= \frac{\pi (l_{-}^2-l_{+}^2)}{G} + \frac{\pi  \left[l_{+}^4 \left(4 \bar{\sigma} ^2 l_{-}^2+1\right)+l_{+}^2 \left(d+4 \bar{\sigma}
^2 l_{-}^4-2 l_{-}^2\right)-l_{-}^2 d+l_{-}^4\right]}{2 G d},
\end{equation}
which also corresponds to the CDL case, now for upward tunneling.

\section{Numerical Analysis of upward/downward Tunneling }\label{3}
	
\begin{figure}
\centering
  \includegraphics[width=75mm]{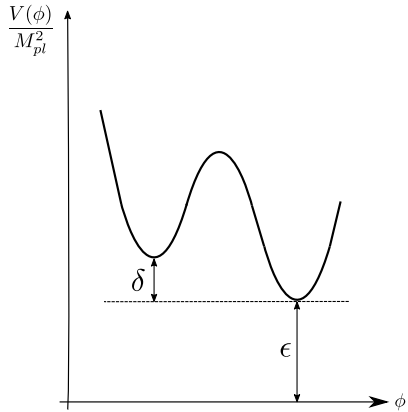}
  \caption{The potential of the tunneling configuration showing the definition of the parameters $\delta$ and $\epsilon$.}
  \label{fig00}
\end{figure}

We compare the tunneling rates of the SdS-SdS upward/downward phase transitions with an arbitrary cosmological constant by performing a full numerical analysis. The parameter $\delta$ represents the difference between the vacua while $\epsilon$ is measuring the upward shift of the potential as shown in Fig. \ref{fig00}. These parameters are related to the cosmological constants in each vacuum via, \footnote{
The units of the quantities are $c=\hbar =1, [\sigma]= M_{pl}, [\epsilon]= [\delta]= M_{pl}^2, [G M_+]=l_{pl}$.}
\begin{align}\label{eq:19}
&\Lambda_+= \epsilon + \delta= \frac{3}{l_+^2}, & \Lambda_-= \epsilon = \frac{3}{l_-^2}.
\end{align}
We explore the range $\epsilon = 10^{-7}  M_{pl}^2$ to $\epsilon = 5 \times 10^{-6} M_{pl}^2$, which corresponds to $l _+ \approx 655l_{pl}$ to $l _+ \approx1195 l_{p l}$  and $l_-\approx775 l_{p l}$ to  $l_-\approx 5477  l_{p l}$, while we define $p$ and $q$ as the fraction of the black hole mass with respect to the Nariai mass, \footnote{ The limit where the two horizons coincide and  $\frac{(GM)^2}{l^2}\rightarrow \frac{1}{27}$,  is known as the Nariai space-time.}
\begin{align}\label{eq:20}
&p= \frac{G M_{+}}{G M_{N_+}} =  \frac{G M_{+} \sqrt{27}}{l_+}, & q= \frac{G M_{-}}{G M_{N_-}} = \frac{G M_{-} \sqrt{27}}{l_-}.
\end{align}

Given these values of the parameters, we explore how the shifting of the potential, $\epsilon$, the difference between the vacua, $\delta$, as well as the tension, $\sigma$, affect the tunneling process. First, we solve (\ref{56}) for the radius of the bubble, $R$, and (\ref{eq:9}) for the time evolution $\dot{\tau}$. We compute numerically the ratio of the bounce (\ref{209}) to its zero-mass limit (\ref{eq:500}) for downward tunneling and similarly the ratio of (\ref{219}) to (\ref{919}) for upward tunneling, to determine if each process is enhanced or suppressed with the inclusion of the black hole. In other words, we calculate $B$ and by comparing it to $B_{M_{\pm}=0}$ we determine if the tunneling is lower or faster compared to the CDL case. For instance, $B<B_{M_{\pm}=0}$ corresponds to faster tunneling.

Numerical analyses of this type have been performed in the past \cite{Gregory:2013hja,Burda:2015yfa}. In \cite{Gregory:2013hja}, the case of $\epsilon=0$ for downward tunneling was considered for nonzero masses inside and outside the bubble, while in \cite{Burda:2015yfa} the case of nonzero epsilon was studied, for downward tunneling. In both papers, the parameter space of masses was searched numerically to find the preferred tunneling process compared to the CDL scenario and was found that the initial inhomogeneities speed up the tunneling process and depending on the values of the masses the dominant process could be either the tunneling to a/no BH.  In this work, we also perform a numerical search but now over the parameter space in epsilon and delta which gives us the opportunity to better isolate the effects of the cosmological constant in the tunneling process. This is quite interesting not so much for the tunneling downward the potential case as for the tunneling upward one, given that we are interested to answer a question about primordial inhomogeneities and if they can speed up the tunneling up process. This, for example, could help us better understand the initial conditions for inflation.

 We consider two cases. The first one corresponds to the initial and the final state in the same Hubble volume (true vacuum bubble within a false vacuum) while the second one corresponds to the initial and final state being separated by a cosmological horizon (false vacuum bubble inside the true vacuum). Additionally, for the processes that involve the same or a larger mass of the black hole inside the bubble, as compared to the black hole outside the bubble, the evolution of the time coordinate is positive, meaning that time is increasing along the wall trajectory and the bubble is expanding. On the other hand, when the black hole mass is smaller inside the bubble, there is a sign change of $\dot{\tau}$, indicating that after its formation, the bubble is initially contracting and then expanding. 
 
\begin{figure}[!tbp]
  \begin{subfigure}[b]{0.475\textwidth}
    \includegraphics[width=\textwidth]{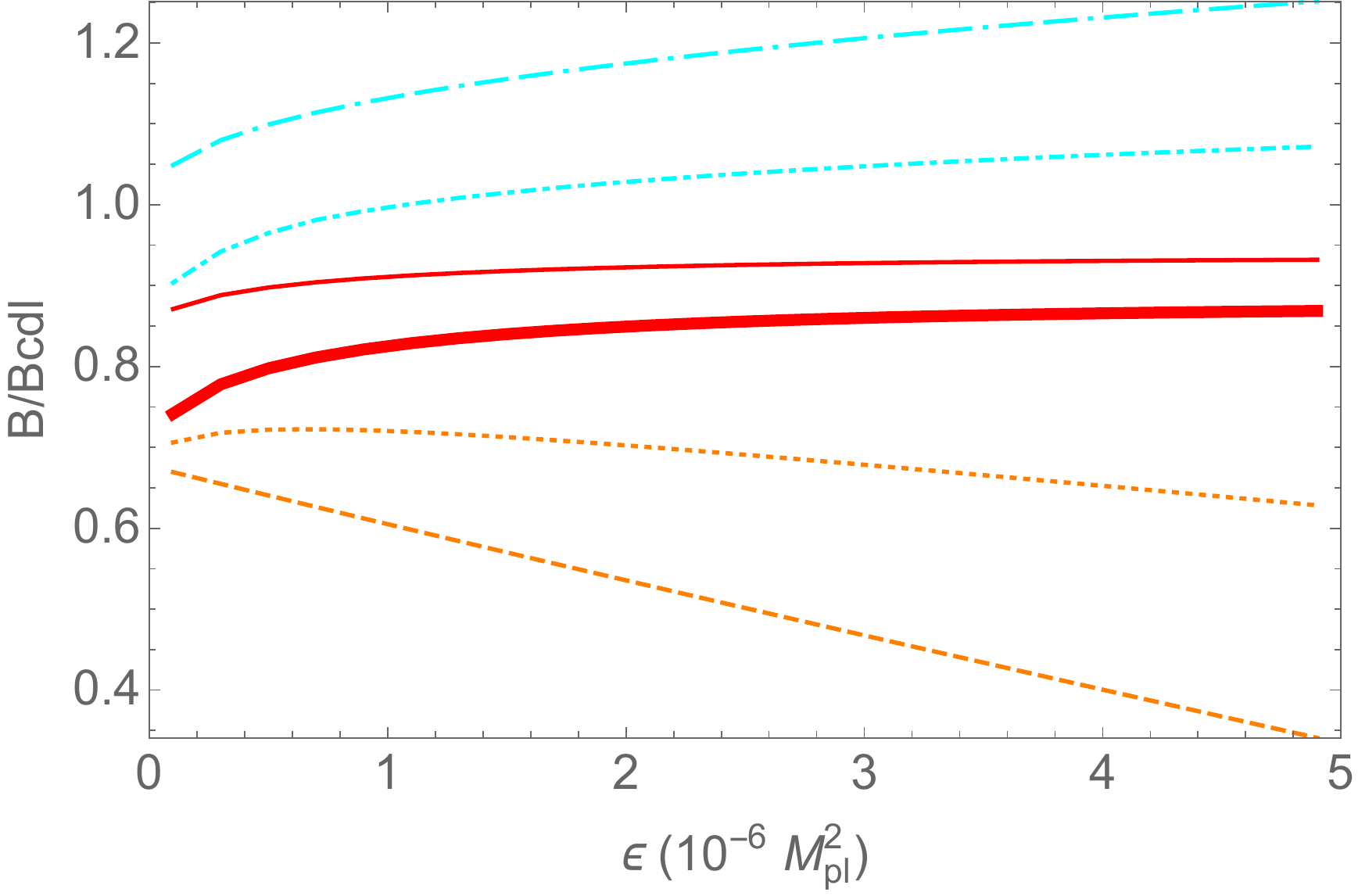}
    \caption{Tunneling downward in the potential.}
    \label{fig:f1}
  \end{subfigure}
  \hfill
  \begin{subfigure}[b]{0.475\textwidth}
    \includegraphics[width=\textwidth]{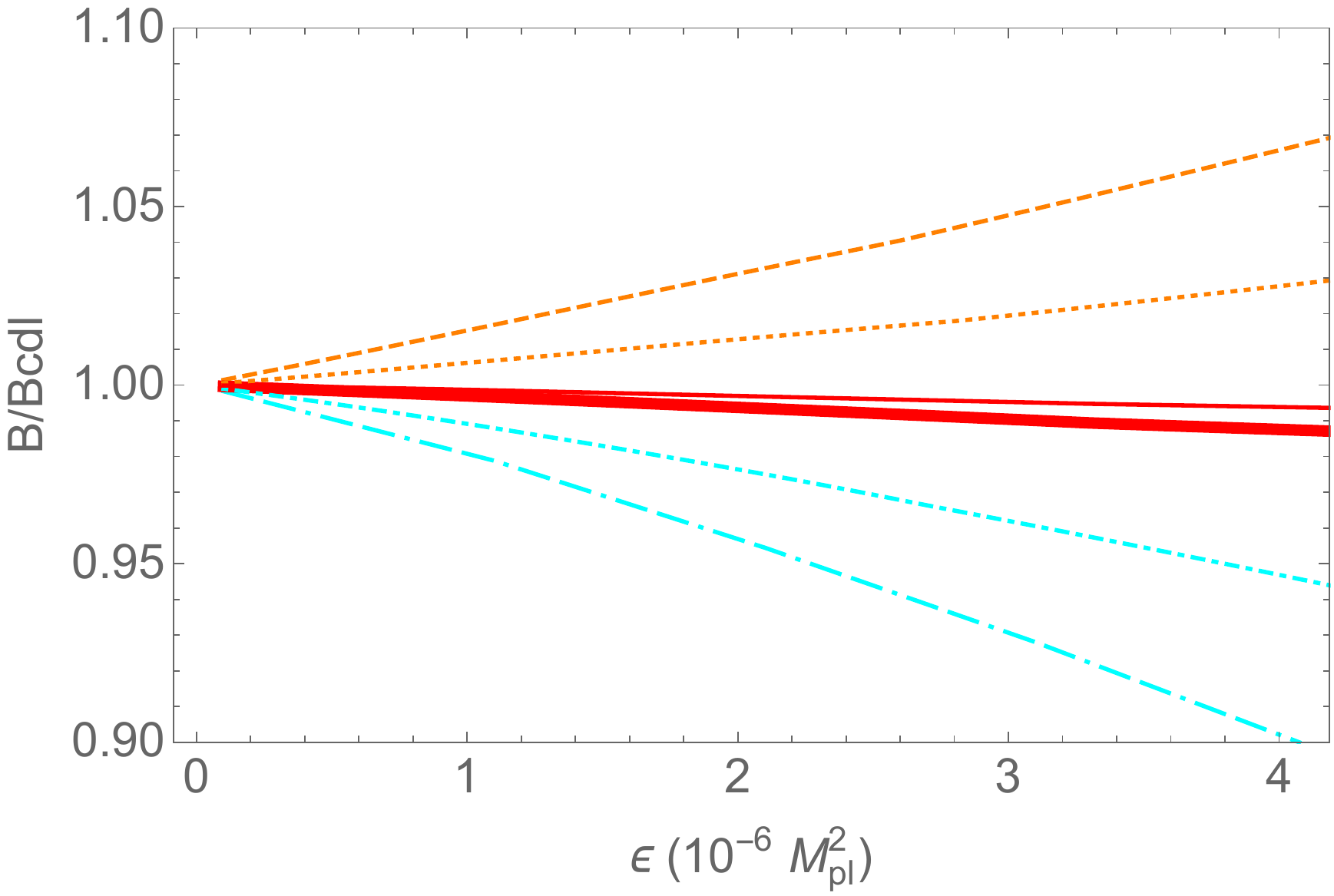}
    \caption{Tunneling upward in the potential.}
    \label{fig:f2}
  \end{subfigure}
\caption{\small The effect of shifting the potential on the tunneling.  On the left picture, for $\sigma=2 \times 10^{-4} M_{pl}$ and $\delta =2 \times 10^{-6} M_{pl}^2$, the double-dashed cyan line represents tunneling from a dS exterior to a SdS interior  of $GM_- \approx 24 l_{p l}$ while the cyan single-dashed line is the tunneling from a SdS exterior of $GM_+ \approx 12 l_{p l}$  to a SdS interior of $GM_- \approx 24 l_{p l}$. The thick red line represents tunneling between SdS vacua of the same black hole mass  $GM_{+}=GM_{-} \approx 24 l_{p l}$  while the thinner line represents the same but for $GM_{+}=GM_{-} \approx 12 l_{p l}$. Finally, the orange dashed line represents nucleation of a dS bubble into a SdS exterior of $GM_+ \approx 24 l_{p l}$ while the dotted orange line is the tunneling from a SdS exterior $GM_+ \approx 24 l_{p l}$ to a SdS interior of $GM_-\approx 12 l_{p l}$. On the right picture, the situation is reversed as, for instance , the double-dashed cyan line represents the nucleation of a false vacuum bubble of a dS interior in a  SdS  of $GM_- \approx 24 l_{p l}$, the single-dashed cyan line is the tunneling from $GM_- \approx 24 l_{p l}$ to $GM_+ \approx 12 l_{p l}$ , etc.}
\label{fig:2}
\end{figure}

A comparison of tunneling rates as a function of $\epsilon$ is shown in Fig.(\ref{fig:2})\footnote{The values of the masses are chosen to be small for numerical convenience. The results could be scaled according to (\ref{eq:19}), since the relevant quantity is the ratio and how it is changing by fixing the mass. This, in any case, does not evade the semiclassical approximation used in the paper, and the values should be considered as part of the proof of concept regarding tiny primordial black holes.}. In Fig.(\ref{fig:f1}), we notice that the fastest phase transition (dashed orange line) represents the nucleation of a dS bubble into a SdS exterior of $GM_+ \approx 24 l_{p l}$, leaving behind the transition that corresponds to tunneling from a SdS exterior $GM_+ \approx 24 l_{p l}$ to a SdS interior of $GM_-\approx 12 l_{p l}$ (dotted orange line). This indicates that the system ``prefers'' to get rid of the black hole altogether in the new vacuum state. The slowest one corresponds to the opposite case, i.e., tunneling from a dS exterior to a SdS interior of $GM_- \approx 24 l_{p l}$ (double-dashed cyan line). We should mention that for these parameters the latter process is subdominant to the CDL case for all values of $\epsilon$. The second slowest rate corresponds to tunneling from $GM_+ \approx 12 l_{p l}$ to $GM_- \approx 24 l_{p l}$ where the mass of the black hole in the new state grows compared to the mass of the old state. Hence, it is natural to expect that the intermediate tunneling rate would correspond to tunneling between SdS vacua of the same black hole mass $GM_{+}=GM_{-} \approx 24 l_{p l}$ (thick red line) or $GM_{+}=GM_{-} \approx 12 l_{p l}$ (thin red line) as is found. 
\begin{figure}[!tbp] 
  \begin{subfigure}[b]{0.475\textwidth}
    \includegraphics[width=\textwidth]{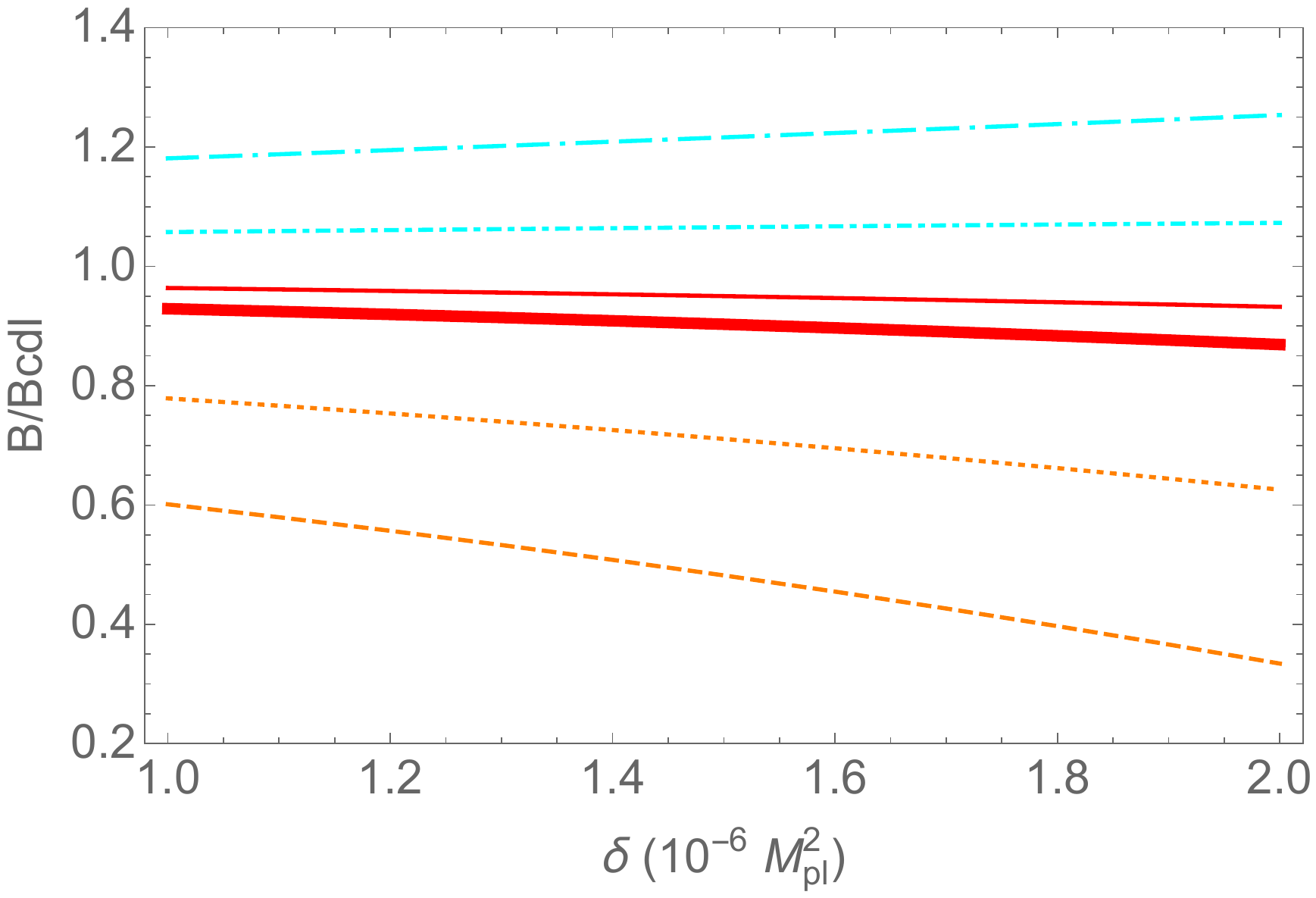}
    \caption{Tunneling downward in the potential.}
    \label{fig:f3}
  \end{subfigure}
  \hfill
  \begin{subfigure}[b]{0.475\textwidth}
    \includegraphics[width=\textwidth]{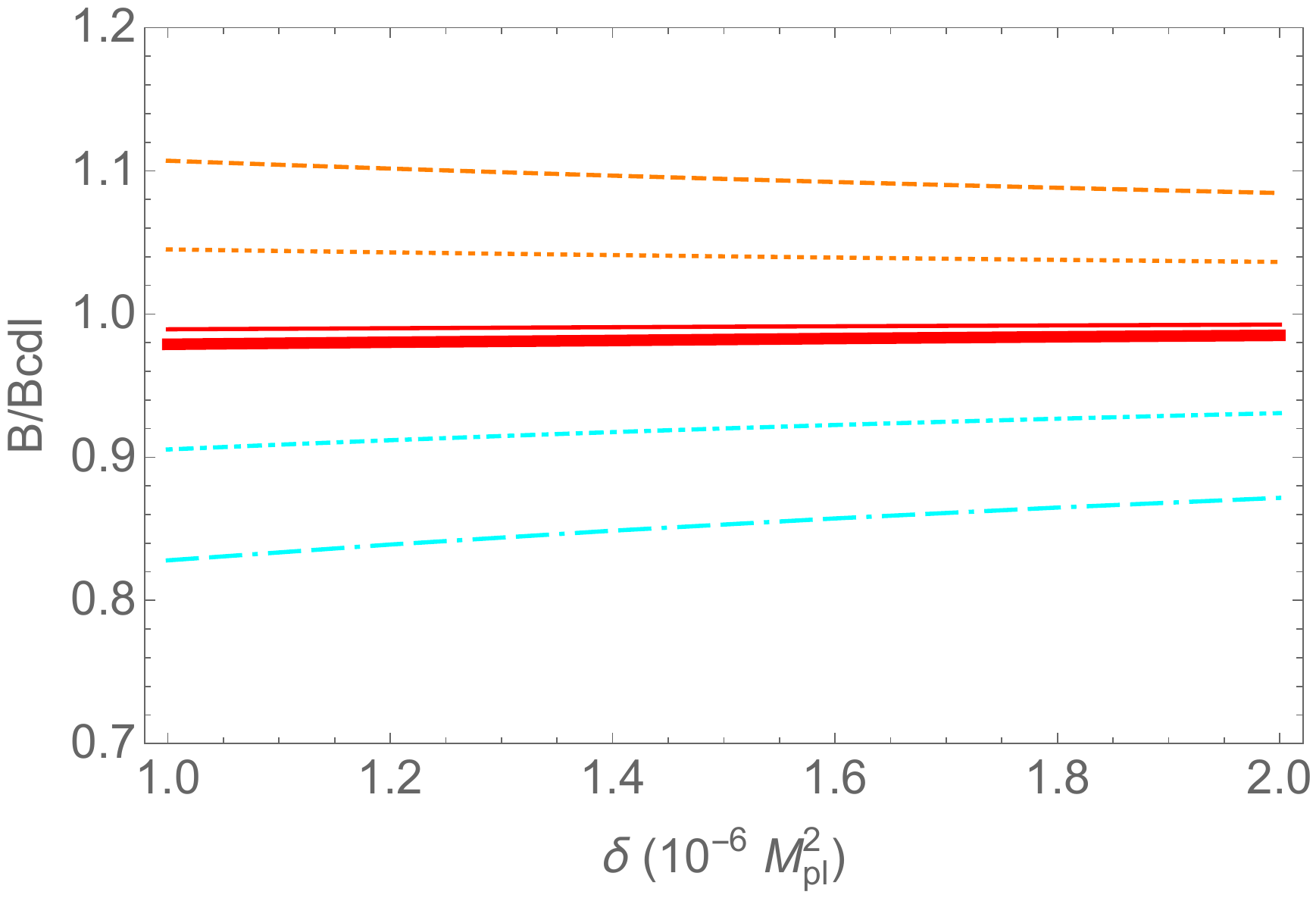}
    \caption{Tunneling upward in the potential.}
    \label{fig:f4}
  \end{subfigure}
  \caption{\small The effect of the difference between the vacua, $\delta$, on the tunneling. We choose $\epsilon=5 \times 10^{-6} M_{pl}^2$ and the rest of the parameters and curves are the same as in Fig.(\ref{fig:2}). In Fig.(\ref{fig:f4}), the orange curves decrease as a function of $\delta$ while the cyan increase, unlike in Fig.(\ref{fig:f2}), where the reverse happens. }
  \label{fig:3}
\end{figure}
 \begin{figure*}
        \centering
        \begin{subfigure}[b]{0.475\textwidth}
            \centering
            \includegraphics[width=\textwidth]{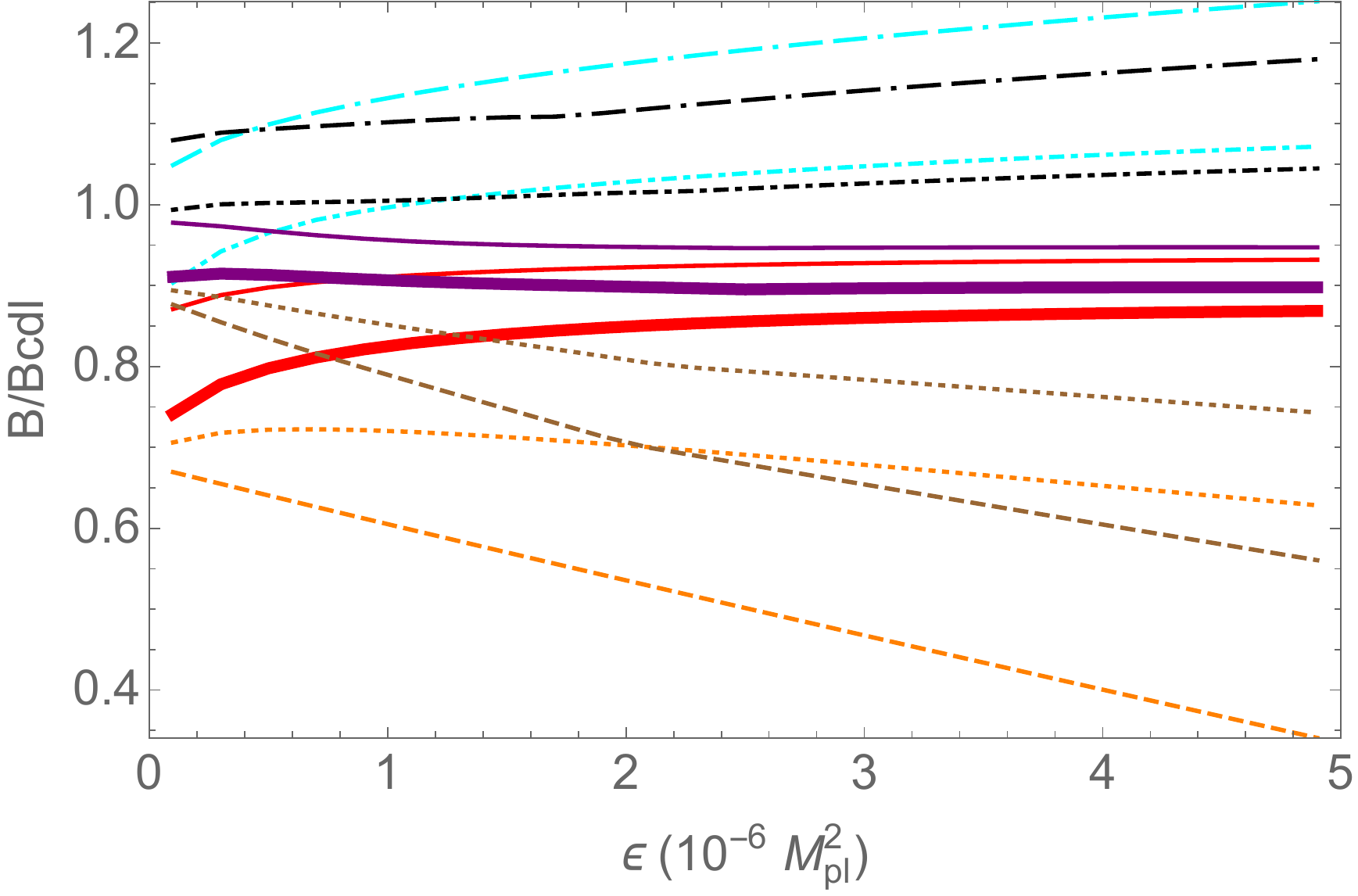}
            \caption[Network2]%
            {{\small Tunneling downward in the potential}}    
            \label{fig4a}
        \end{subfigure}
        \hfill
        \begin{subfigure}[b]{0.475\textwidth}  
            \centering 
            \includegraphics[width=\textwidth]{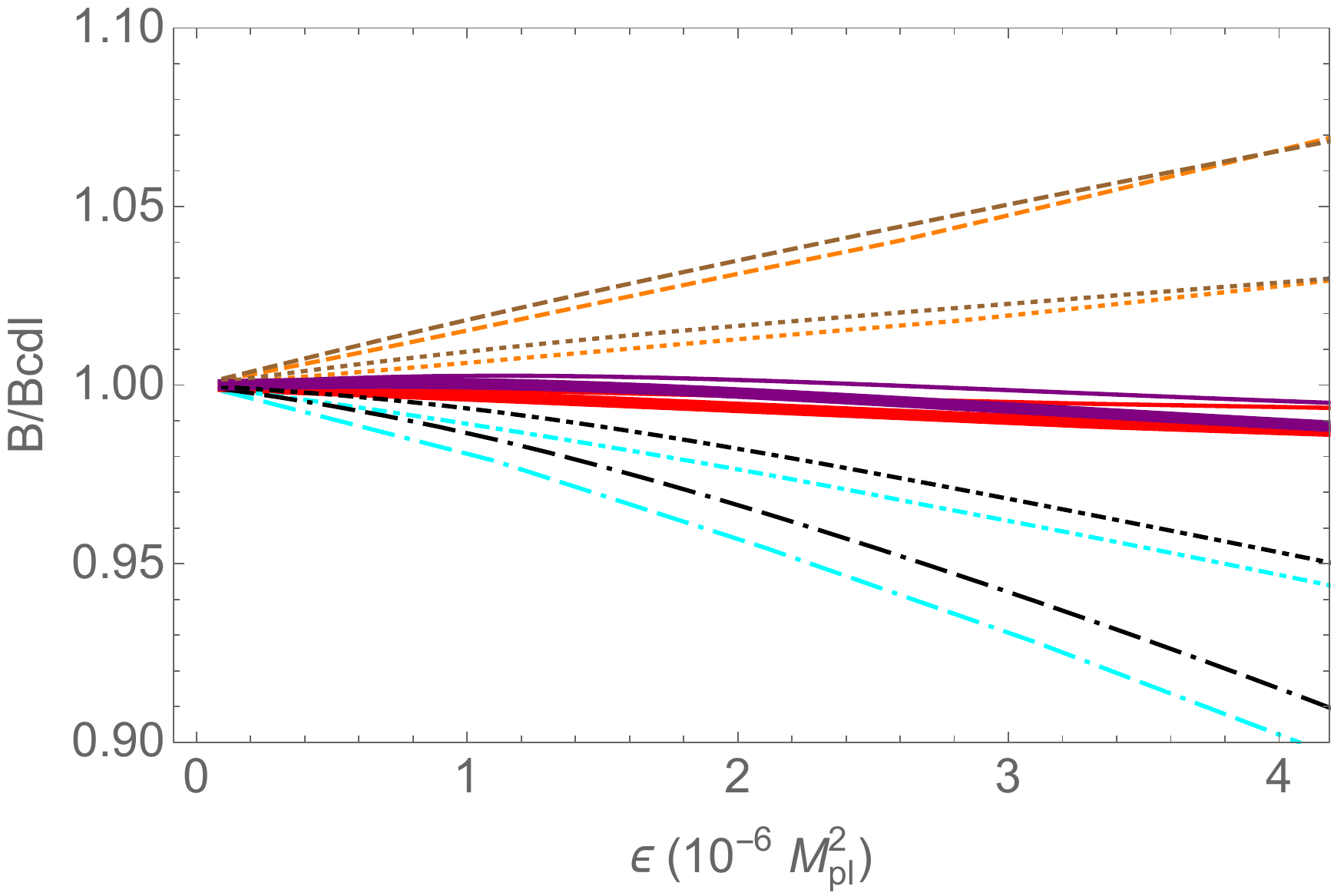}
            \caption[]%
            {{\small Tunneling upward in the potential}}    
            \label{fig4b}
        \end{subfigure}
        \vskip\baselineskip
        \begin{subfigure}[b]{0.475\textwidth}   
            \centering 
            \includegraphics[width=\textwidth]{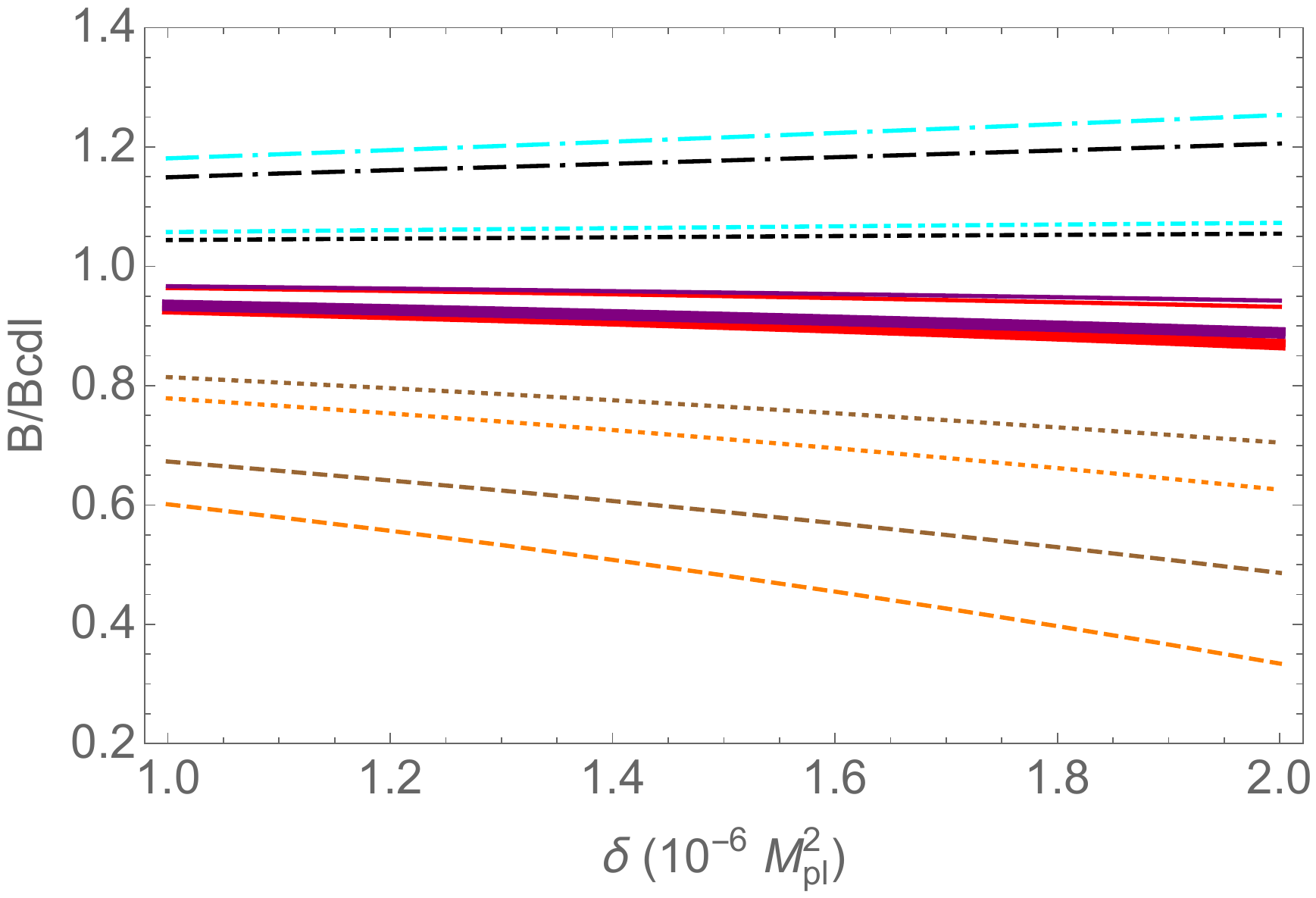}
            \caption[]%
            {{\small Tunneling downward in the potential}}    
            \label{fig4c}
        \end{subfigure}
        \quad
        \begin{subfigure}[b]{0.475\textwidth}   
            \centering 
            \includegraphics[width=\textwidth]{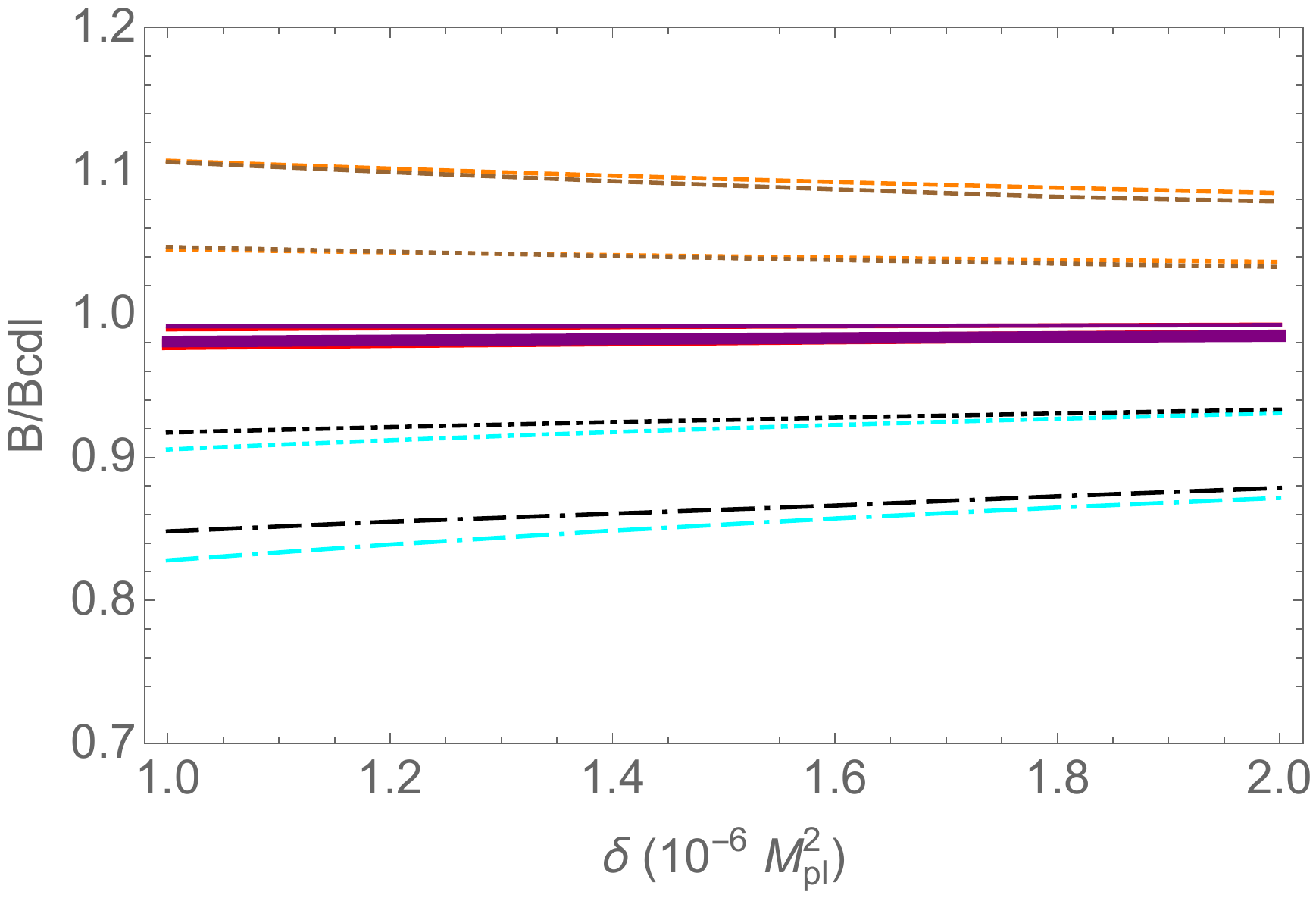}
            \caption[]%
            {{\small Tunneling upward in the potential}}    
            \label{fig4d}
        \end{subfigure}
        \caption[  ]
 {\small The effect of tension on the tunneling as a function of the potential parameters $\epsilon$, $\delta$ and for different black hole masses in the initial and final states. The ranges of the parameters are the same as in Fig.(\ref{fig:2}) and Fig.(\ref{fig:3}). The black, purple and brown lines correspond to $\sigma$=$25 \times 10^{-5} M_{pl}$. The double-dashed black line represents tunneling from a dS exterior to a SdS interior of  $GM_- \approx 24 l_{p l}$,  The dashed brown line represents  tunneling from a SdS exterior of of $GM_+ \approx 24 l_{p l}$ to a dS interior while the purple line is tunneling between  SdS vacua of the same mass $GM_{+}=GM_{-} \approx 24 l_{p l}$. The dashed black line represents a transition from a SdS of $GM_+ \approx 12 l_{p l}$  to a SdS of $GM_- \approx 24 l_{p l}$, the dotted brown line tunneling from a SdS of $GM_+ \approx 24 l_{p l}$ to a SdS of $GM_- \approx 12 l_{p l}$ and the thin purple line tunneling between vacua of the same mass $GM_{+}=GM_{-} \approx 12 l_{p l}$.} 
\label{fig:45}
\end{figure*}

On the other hand, tunneling upward in the potential, or the nucleation of a false vacuum bubble inside the true vacuum as a function of $\epsilon$ is shown in Fig.(\ref{fig:f2}). Here we notice that the fastest tunneling rate corresponds to tunneling from a SdS exterior of $GM_- \approx 24 l_{p l}$ to a dS interior of a higher value of the cosmological constant (cyan double dashed line) while the two lowest tunneling rates are subdominant to the CDL case (dashed and dotted orange lines). Thus, we observe that even in the false bubble nucleation, as the cosmological constant takes on higher values, the inhomogeneities (black holes) are more likely to vanish. Even though the upward tunneling rate is largely suppressed compared to the  downward tunneling rate, we notice that the fastest rate takes on up to a 10 percent enhancement compared to the CDL one, over the range of parameters explored. In both cases, it is evident that as the value of $\epsilon$ grows, the tunneling to lower values of the mass is enhanced while the tunneling to larger masses is suppressed compared to the CDL case.

Next, we study how the difference between the vacua affects the tunneling rate(see Fig.(\ref{fig:3})).  For tunneling downward the potential, as the value of $\delta$ increases, the largest tunneling rate corresponds to tunneling from a SdS exterior of $GM_- \approx 24 l_{p l}$ to a dS interior (Fig.(\ref{fig:f3})). For tunneling upward in the potential (Fig. (\ref{fig:f4})), the fastest tunneling rates (double dashed and single dashed line) decrease as the difference between the vacua increases. To have the most enhanced rate, we need small values of $\delta$ and, again, no black hole in the end state. 

To complete the picture, we briefly comment on the effect of the tension on the tunneling rates.  As the tension increases, the tunneling probability becomes smaller, confirming the expectation that as the walls get thicker, the bubble nucleation becomes less probable as seen in Fig.(\ref{fig:45}). There, the black, purple and brown lines represent tension of $\sigma=25 \times 10^{-5} M_{pl}$ and $\delta =2 \times 10^{-6} M_{pl}^2$  in Fig.(\ref{fig4a}, \ref {fig4b}) while in (\ref {fig4c}, \ref {fig4d}) we have $\epsilon=5 \times 10^{-6} M_{pl}^2$.

The goal of this work is not to exhaustively search the parameter space for the masses when $\epsilon=0$ as done in \cite{Gregory:2013hja,Burda:2015yfa}. Although when taking the limit of $\epsilon=0$ and using the appropriate choice of parameters our results coincide with theirs, the novelty of this paper is to check explicitly  the effect of epsilon and delta on the tunneling between an initial mass and a final mass for a given set of parameters. For every initial mass that is chosen, 3 distinct cases are considered, namely a final mass smaller, equal or larger than the initial mass. The question is which of these three cases is the minimum compared to the CDL case as we vary epsilon and delta. We noticed that as the value of epsilon increases, the preferred process compared to the CDL case is the tunneling to no black hole (zero final mass) for the tunneling downward case. The next step was to check if within this parameter space, as we increase epsilon, the tunneling to no black hole process is still the nonzero one even for the tunneling upward the potential. Although the latter is suppressed compared to the tunneling favored, there is a part of the parameter space (the one presented in the paper) where there is an enhancement in the transition rate compared to the upward CDL one and the black holes act as seeds of bubble nucleation in this case as well. This completes the proof of concept procedure.

\section{FGG mechanism}\label{4}
To understand how inflating regions may be spawned from noninflating ones, it is worthwhile to study processes such as the FGG mechanism. In this case, a false vacuum bubble tunnels through a wormhole to produce an inflating region \cite{Aguirre:2005nt,Farhi:1989yr}. As we take the zero-mass limit of this process, a totally disconnected phase that includes the new vacuum is nucleated while the initial spacetime is maintained. This is in contrast with the CDL scenario where the tunneling from Minkowski to a higher energy density vacuum is prohibited.

To calculate the rate of the FGG mechanism, first, we write down the Euclidean action of the dS to S process,
\begin{equation}
B_{dS/S}=I_{E}-I_{dS}= -\frac{A_{-}}{4 G}  +\frac{1}{4 G} \int d\lambda [(2 R f_{+}-R^2  f'_{+})\dot{\tau}_{+}-(2 R f_{-}-R^2  f'_{-})\dot{\tau}_{-}],
\end{equation}
where
\begin{equation}
I_{dS}=- \frac{A_{c_+}}{4 G}.
\end{equation}
By using (\ref{eq:100}) and
\begin{align}
&I_f=-\frac{A_{c_{+}}}{4 G},& I_t=- \frac{A_{-}}{4 G},
\end{align}
we arrive at
\begin{equation}
B_{FGG}=-\frac{A_{c_+}}{4 G} +\frac{A_{-}}{4 G}  +\frac{1}{4 G} \int d\lambda [(2 R f_{+}-R^2  f'_{+})\dot{\tau}_{+}-(2 R f_{-}-R^2  f'_{-})\dot{\tau}_{-}].
\end{equation}
\begin{figure*}
\centering
\includegraphics[width=75mm]{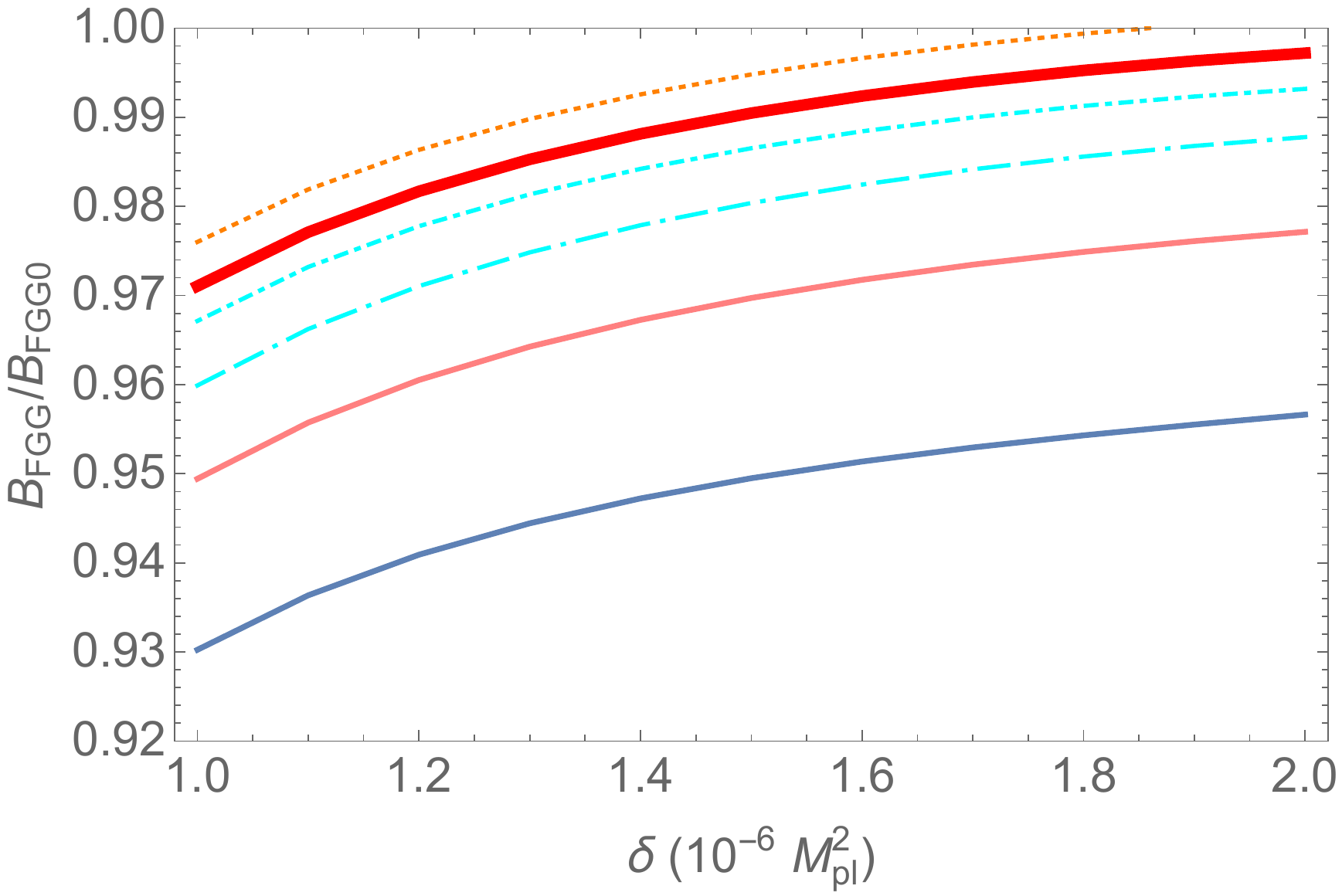}
    \caption{\small FGG mechanism with/without a black hole in the final state. The thin pink line represents tunneling from S with $GM_{+}\approx 48 l_{p l}$ to dS while the grey thin line represents tunneling from S with $GM_{+}\approx 72 l_{p l}$. Along with the cyan double dashed line (tunneling from S with $GM_{+}\approx 24 l_{p l}$ to dS) these represent the S to dS processes. The rest of the lines represent the S-SdS processes. The single-dashed cyan line is the tunneling from $GM_- \approx 24 l_{p l}$ to $GM_+ \approx 12 l_{p l}$, the dotted orange line represents $GM_- \approx 12 l_{p l}$ to $GM_+ \approx 24 l_{p l}$ while the thick red line is $GM_- \approx 24 l_{p l}$ to $GM_+ \approx 24 l_{p l}$.  }
    \label{fig:f7}
\end{figure*}
The zero mass limit of this process leads to
\begin{align}\label{eq:309}
B_{FGG_0}&= -\frac{A_{c_+}}{4 G} +\frac{1}{4 G} \int d\lambda [(2 R f_{+}-R^2  f'_{+})\dot{\tau}_{+}-(2 R f_{-}-R^2  f'_{-})\dot{\tau}_{-}] \nonumber\\ &=- \frac{ \pi l^2}{G} \frac{1+ 8 \bar{\sigma}^2 l^2}{ (1+ 4 \bar{\sigma}^2 l^2)^2}.
\end{align}
The minus sign in (\ref{eq:309}) is related to a sign choice we are forced to make due to quantum cosmological boundary conditions so as to keep the transition probability smaller than one.

Similar to the previous section, we perform a numerical analysis on the FGG mechanism, by considering two distinct cases. In Fig.(\ref{fig:f7}), among the S-dS processes (pink, grey and blue dashed lines), the  tunneling from S with $GM_{+}\approx 72 l_{p l}$ to dS dominates while, among the S-SdS processes we conclude that the dominant process is the one that tunnels to a smaller mass black hole (single-dashed cyan line). Overall the preferred state is the one that tunnels to no black hole. It is evident that the tunneling to no black hole is favored not only with the inclusion of a positive cosmological constant in the initial phase but in the FGG mechanism as well. 

This gives us the opportunity to make a relative comparison of the FGG mechanism\footnote{Note that this does not correspond to the $\epsilon\rightarrow 0$ limit of the processes described in Section \ref{3}, even though $\epsilon=0$ is valid in this case. In fact, as explained above, this limit is not allowed in the CDL scenario.} with the up-tunneling process described in Section $\ref{3}$. Since $B_{FGG} \backsim B_{FGG_0}$, the inclusion of a black hole in the initial phase does not have a big effect on the FGG tunneling rate while in Fig.(\ref{fig:f4}), especially for the tunneling to no black hole (cyan single dashed and double dashed lines), clearly $B<B_{CDL}$. This shows that the inclusion of a black hole in the initial phase makes the nonzero $\epsilon$ case more sensitive to transition to a dS phase than the FGG mechanism.

\section{Conclusion}\label{5}
In this work, we have explored the nucleation of true and false vacuum bubbles. Our discussion was restricted to a positive cosmological constant, including a black hole in both the initial and final states. By separating the difference between the vacua, $\delta$, and the vertical shift of the potential, $\epsilon$, we study the tunneling probability of the processes. We find that as the potential shifts to higher values of the cosmological constant, the nucleation rate of true and false vacuum bubbles is enhanced compared to the CDL rate. Overall, we explored different values of the black hole masses in both vacua and we found, as a proof of concept,  that the fastest tunneling rate (for all cases) corresponds to an end state of no black hole. Especially for the nucleation of false vacuum bubbles, it means that the tunneling to higher values of the cosmological constant tends to remove inhomogeneities. This could have important consequences for the early universe, for example, this could be a process from which the initial inhomogeneities in a noninflating universe vanish. Furthermore, we explored the effect of the difference between the vacua on the tunneling rate. As in the case of $\epsilon$, we find that the no-black hole end state leads to the most enhanced rate.

The creation of inflating regions out of non inflating ones was analyzed in the context of the FGG mechanism. Two cases were considered, the first one being the nucleation of a dS bubble, completely disconnected from the initial Schwarzschild spacetime. We notice that within this tunneling process, the production of a false vacuum bubble becomes more likely. In the second case, we allow for a remnant black hole in the final state. Comparing all the tunneling events in Fig. (\ref{fig:f7}), we conclude that the final state without the inclusion of a black hole is slightly favored.

This provides a new way to make a relative comparison between the FGG mechanism and the tunneling upward in the potential with a nonzero cosmological constant in both vacua. While for both processes the most likely scenario is the complete elimination of inhomogeneities, we observe that for the same range of $\delta$, the nonzero value of $\epsilon$ is essential in speeding up the tunneling process. This indicates that not only the inclusion of a black hole is necessary in the initial phase to enhance the tunneling rate, but the inclusion of a cosmological constant as well. While this comparison cannot exclude the FGG mechanism as a physical process, at least within the parameter range explored in this paper, it shows that the existence of a nonzero value of $\epsilon$ can enhance the elimination of inhomogeneities in the early universe, thus providing a sufficiently smooth patch for the onset of inflation. Further exploration of the parameter space would be necessary to make these arguments more concrete. 

In terms of understanding the initial conditions for inflation, as well as the mechanisms that lead to transitions between vacua, it is important to consider all the allowed tunneling scenarios and their likelihood as this will help understand the preferred transitions among the vacua in the string theory landscape. In this work, we have used the Euclidean instanton approach to explore all the allowed transitions with a nonzero positive cosmological constant. We have extended the analysis to include tunneling downward in the potential as well as the FGG mechanism. It remains of interest to use our method to explore the formation of AdS bubbles as this could be deeply linked to the information loss problem or to the study of the stability of the Higgs vacuum since these nucleation seeds could drastically alter  the time it takes to decay to a different standard model.

\acknowledgments
I would like to thank Claudia de Rham and Glenn Starkman for useful discussions at the beginning of this project, as well as Craig Copi, Tate Deskins, Shamreen Iram, Laura Johnson and Soumyajit Bose for useful comments on the manuscript. Finally, I would like to thank the referee for his constructive suggestions and the discussion about the Hawking-Moss instanton.

\appendix

\section{Conical angles}\label{6}

Here we present a more pedagogical way of addressing the issue of conical singularities discussed in \cite{Gregory:2013hja}. For a spherically symmetric metric we have,
\begin{equation}
ds^2= f(r) d{\tau}^2 + \frac{1}{f(r)}d r^2 + r^2 d\Omega^2,
\end{equation}
Expanding around one of the horizons, $ r=r_i $, we write
\begin{align}
f(r)&=f(r=r_i) + f'(r=r_i) (r-r_i).
\end{align}
Further, on the metric, we perform the transformation $ d \rho = \frac{dr}{\sqrt{f(r)}} $. Integrating this expression we find 
 \begin{align}
& \rho=\frac{1}{\sqrt{f'(r=r_i)}} 2 \sqrt{r-r_i}, & r = \frac{f'(r=r_i)}{4} \left(\rho^2+ \frac{4 r_i}{f'(r=r_i)}\right), &  f(r) = \frac{f'(r=r_i)^2 \rho^2}{4} .
 \end{align}
The Euclidean time is periodic,  $\tau=\frac{\phi \beta}{2 \pi}$ and $0 \leq \tau \leq \beta$. Combining everything together we arrive at, the transformed metric,
\begin{equation}
ds^2=\frac{f'(r=l)^2 \beta^2}{16 \pi^2} \rho^2d{\phi}^2 + d \rho^2 + r(\rho)^2 d\Omega^2.
\end{equation}
This has the form of a cone
\begin{equation}
ds_{cone}^2= \alpha^2 \rho^2 d{\phi}^2 + d \rho^2,
\end{equation}
where $\alpha = 1- \frac{\Delta}{2 \pi}$. 
When $\alpha=1$, the deficit angle is 0 which implies that $\frac{f'(r=r_i)^2 \beta^2}{16 \pi^2} =1$.  There are metrics which always have a deficit on one of the horizons. Let us consider for example, the Schwarzschild-de Sitter metric,
\begin{equation}
ds^2=\frac {(\frac{2 G M}{r^2}-\frac{2 r}{l^2})^2 \beta^2 \rho^2}{16 \pi^2} d{\phi}^2 + d \rho^2 + r(\rho)^2 d\Omega^2.
\end{equation}
In a SdS spacetime, the black hole horizon and the cosmological horizon are located at \cite{Shankaranarayanan:2003ya},
\begin{equation}
r_{h_{\pm}}=\frac{2 l_{\pm} }{\sqrt{3}}  \cos\left[\frac{\pi }{3}+\frac{1}{3} \cos
   ^{-1}\left(3 \sqrt{3} \frac{M_{\pm}}{l_{\pm}}\right)\right],
\end{equation}
and,
\begin{equation}
r_{c_{\pm}}=\frac{2 l_{\pm} }{\sqrt{3}}  \cos \left[\frac{\pi }{3}-\frac{1}{3} \cos
   ^{-1}\left(3 \sqrt{3} \frac{M_{\pm}}{l_{\pm}}\right)\right].
\end{equation}
From a physical point of view, the periodicity $\beta$ is equal to the inverse of the temperature and the cases to consider are,
\begin{itemize}
\item If $\beta_c=\frac{1}{T_c} $ and $r=r_c$   then $T_c= \frac{1}{4 \pi} (\frac{2 G M}{r_c^2}-\frac{2 r_c}{l^2})$.
\item If $\beta_h=\frac{1}{T_h} $ and $r=r_h$   then $T_h= \frac{1}{4 \pi} (\frac{2 G M}{r_h^2}-\frac{2 r_h}{l^2})$.
\item Finally we can have the case where $\beta=\frac{1}{T} $ allowing for 2 deficit angles at $r=r_c$ and  $r=r_h$.
\end{itemize}

\newpage
\bibliographystyle{apa}
\bibliography{Tunneling_of_Black_holes_in_spacetimes_of_arbitrary_cosmological_constant_prd_v3}
\end{document}